\documentclass[12pt, draftcls, peerreview, onecolumn]{IEEEtran}
\usepackage{times}
\usepackage{amsmath}
\usepackage{amssymb}
\usepackage{amsthm}
\usepackage{verbatim}
\usepackage{enumerate}
\usepackage{graphicx}
\usepackage{algorithm}
\usepackage{algpseudocode}
\usepackage{cite}
\usepackage{hyperref}

\newtheorem*{definition*}{Definition}

\newtheorem{example}{Example}

\begin{document}
\title{On Efficient Decoding and Design of Sparse Random Linear Network Codes}

\author{Ye~Li,~Wai-Yip~Chan,~Steven~D.~Blostein  
\thanks{The authors are with Department of Electrical and Computer Engineering, Queen's University, Kingston, Canada (email: \{y.li, chan, steven.blostein\}@queensu.ca)}}

\IEEEoverridecommandlockouts

\setcounter{page}{1}

\maketitle
%\IEEEpeerreviewmaketitle
\begin{abstract}
Random linear network coding (RLNC) in theory achieves the max-flow capacity of multicast networks, at the cost of high decoding complexity.  To improve the performance-complexity tradeoff, we consider the design of sparse network codes.  A generation-based strategy is employed in which source packets are grouped into overlapping subsets called generations. RLNC is performed only amongst packets belonging to the same generation throughout the network so that sparseness can be maintained.  In this paper, generation-based network codes with low reception overheads and decoding costs are designed for transmitting of the order of $10^2$-$10^3$ source packets. A low-complexity overhead-optimized decoder is proposed that exploits ``overlaps'' between generations. The sparseness of the codes is exploited through local processing and multiple rounds of pivoting of the decoding matrix. To demonstrate the efficacy of our approach, codes comprising a binary precode, random overlapping generations, and binary RLNC are designed. The results show that our designs can achieve negligible code overheads at low decoding costs, and outperform existing network codes that use the generation based strategy.

\end{abstract}
\thispagestyle{headings}
\section{Introduction}
Random linear network coding (RLNC) in theory achieves the max-flow capacity of a multicast network\cite{Ahlswede2000}, \cite{Ho2006} but unfortunately has high decoding complexity. The decoding requires solution of a general system of linear equations in $M$ unknowns to recover $M$ source packets, resulting in high $\mathcal{O}(M^3)$ computational cost using Gaussian elimination (GE).

To reduce decoding cost, a possible solution is to group source packets into subsets called \textit{generations}. By performing RLNC only among packets belonging to the same generation \cite{Chou2003}, a sparse system is obtained. In \cite{Maymounkov2006}, it is established that intermediate nodes can randomly \textit{schedule} generations to perform coding without acknowledgment between nodes. The scheme is referred to as generation-based network coding (GNC) and is the focus of this paper. 

A complete GNC system consists of a \textit{code}, \textit{scheduling strategy}, and \textit{decoder}. The GNC code specifies how generations are formed and how coded packets are generated at the source node. The scheduling strategy at intermediate nodes determines from which generation to re-encode a packet when there is a transmission opportunity. At each destination, the GNC decoder recovers source packets from received coded packets of different generations.

A key GNC performance metric is reception \textit{overhead}, defined as the excess in received packets over the number of source packets needed to decode. The overhead can be classified into \textit{encoder-induced}, \textit{decoder-induced}, and \textit{network-induced}. Encoder-induced overhead (or \textit{code overhead} for brevity) is introduced if there exists at least one subset with size $k\leq M$ of the encoded packets that is not a linearly independent set; decoder-induced overhead is incurred if the decoder is unable to decode when $M$ linearly independent packets are received; network-induced overhead is incurred if scheduling and re-encoding at intermediate nodes introduces linear dependency.

We focus on encoder and decoder-induced overhead. In \cite{Thibault2008, Silva2009a, Li2011a, liye2012, Tang2012, Mahdaviani2012}, GNC codes with \textit{generation-by-generation} (G-by-G) decoding are designed. The G-by-G decoder decodes within generations and decoded packets are subtracted from the received packets of other generations that overlap with the decoded generations. G-by-G decoding can result in a high decoder-induced overhead.

In this paper, we design GNC codes for moderate size $M$ that have low code overhead and a decoder with zero decoder-induced overhead. Moderate refers to when $M$ is of the order of $100\mathrm{s}$-$1000\mathrm{s}$, e.g., as seen in streaming media. We first propose a low-complexity overhead-optimized GNC decoder which succeeds as soon as $M$ linearly independent packets are received and hence has zero decoder-induced overhead. While overhead-optimized decoding has high decoding cost for large $M$, we show that for moderate $M$ it may bring considerable advantages. The proposed decoder, termed \textit{overlap-aware} (OA), exploits the sparseness of GNC codes.% and provides flexibility to control the trade-off between overhead and decoding cost.

The OA decoding first processes received packets \textit{locally} within generations and then pivots the \textit{global} sparse linear system to lower the computational cost. While pivoting has been employed in the decoding of sparse erasure-correction codes such as LDPC \cite{Burshtein2004, Pishro-Nik2004, Paolini2012} and raptor codes \cite{ShokrollahiLuby09}, its application to decode GNC codes has yet to be investigated. A crucial aspect of GNC is that GNC-coded packets mix with others from the same generation. Hence, local processing and multiple rounds of pivoting are needed for efficient decoding. 

For the proposed OA decoder, we propose a GNC code that features binary precoding \cite{Shokrollahi2006}\cite{Maymounkov2002}, random overlapping generations \cite{Li2011a}, and binary RLNC. Precoding and overlapping both reduce code overhead. We show that the proposed code can achieve close-to-zero code overhead and efficient OA decoding. 

Similar types of overhead-optimized decoding of GNC codes have been considered previously. In \cite{Heidarzadeh2010, Heidarzadeh2011, Fiandrotti2014, Heide2014}, GNC based on banded matrices are designed. Each band of the decoding matrix corresponds to a generation. The decoding therein uses straightforward GE, which preserves the banded structure during row reductions and therefore has a low decoding cost.

Our work improves upon \cite{Heidarzadeh2010, Heidarzadeh2011, Fiandrotti2014, Heide2014} in several major ways. First, the proposed OA decoder has low complexity for GNC codes with more general overlapping patterns while the straightforward GE decoder in \cite{Heidarzadeh2010, Heidarzadeh2011, Fiandrotti2014, Heide2014} only applies to GNC codes with banded decoding matrices. To achieve close-to-zero overhead, GNC codes with random overlap may be more desirable as we will show. Second, the proposed design creates sparser codes due to the combined use of precoding and random overlap, resulting in lowered computational costs, while no precoding is used in \cite{Heidarzadeh2010, Heidarzadeh2011, Fiandrotti2014, Heide2014}.

The rest of the paper is organized as follows: Section \ref{model} presents the structure of generation-based network codes. We review existing decoders and propose the OA decoder for GNC codes in Section \ref{decoder_desp}. Section \ref{design} presents the code design, its OA decoding, and performance analysis. In Section \ref{section: evaluation}, we evaluate our design by simulation. The conclusion is given in Section \ref{summary}.

\section{Model: Generation-based Network Coding}\label{model}
Consider that $M$ source packets, $\mathcal{S}=\{\mathbf{s}_i,1\leq i\leq M\}$, are to be sent to destination nodes over a network that contains intermediate nodes. Links are lossy and are modeled as erasure channels. Each source packet consists of $K$ source symbols from a finite field $\mathbb{F}_q$, where $q$ is the finite field size. Each $\mathbf{s}_i$ is a $K$-length row vector on $\mathbf{F}_q$. A set of $L$ generations $\mathcal{G}=\{\mathcal{G}_1,\ldots,\mathcal{G}_L\}$ are constructed from the $M$ source packets. Each generation $\mathcal{G}_l=\left\{\mathbf{s}_1^{(l)},\ldots,\mathbf{s}_{G_l}^{(l)}\right\}, l=1,\ldots, L$, is a subset of source packets of size $G_l=|\mathcal{G}_l|$, where $\mathbf{s}_j^{(l)}=\mathbf{s}_i$ for some $1\leq i\leq M$ for each $j=1,\ldots, G_l$. We assume that $\cup_{l=1}^L\mathcal{G}_l=\mathcal{S}$, i.e., each source packet is present in at least one generation. A one-to-one index mapping $f_l(\cdot): j\rightarrow i$ indicates that the $i$-th source packet is selected as the $j$-th packet in $\mathcal{G}_l$. The source packet indices in $\mathcal{G}_l$ are stored as the set $\mathcal{I}_l=\left\{f_l(1),f_l(2),\ldots, f_l(G_l)\right\}$. $\mathcal{G}$ is said to be \textit{disjoint} if $\mathcal{I}_i\cap \mathcal{I}_j=\emptyset, \forall i\neq j$ or else \textit{overlapping}, and is of \textit{equal-size} if $G_i=G_j, \forall i, j$ or else of \textit{unequal-size}. For overlapping $\mathcal{G}$, $\sum_{l=1}^LG_l>M$. We assume that the index mappings are made known to the destination nodes.

A GNC code is defined on $\mathcal{G}$ as follows: for each transmission from the source node, a coded packet is generated as a random linear combination of source packets in a randomly chosen generation; coefficients are chosen from $\mathbb{F}_q$. Let $\mathcal{P}=\{p_1,p_2\ldots,p_L\}$ be the set of probabilities where $p_l$ denotes the probability that generation $\mathcal{G}_l$ is chosen when generating a packet, $1\leq l\leq L$. The GNC code is then characterized by $(\mathcal{G},\mathcal{P},q)$. The GNC codes are \textit{rateless}, meaning that a potentially unlimited number of coded packets may be generated. The source node is informed to stop transmission only after the destinations have recovered all of the source packets.

At intermediate nodes, packets are re-encoded from previously received packets of a chosen generation. Re-encoded packets are assumed to be random linear combinations of the received packets. The procedure for choosing the generation to re-encode for the next transmission is called \textit{scheduling}. Different scheduling strategies may be used, such as random scheduling \cite{Maymounkov2006}, \cite{Li2011a} and maximum local potential innovativeness scheduling \cite{liye2013}. 

At a destination, received packets of $\mathcal{G}_l,l=1,\ldots,L$ are in the form of $\mathbf{r}^{(l)}=\sum_{i=1}^{G_l}g_i^{(l)}\mathbf{s}_i^{(l)}$, where $g_i^{(l)}$ is an encoding coefficient from $\mathbf{F}_q$. We refer to $\mathbf{g}^{(l)}=[g_1^{(l)},g_2^{(l)},\ldots,g_{G_l}^{(l)}]$ as a \textit{generation encoding vector} (GEV) of $\mathcal{G}_l$. The GEV is delivered in the header of each coded packet. Each GEV can be transformed to a length-$M$ \textit{encoding vector} (EV), denoted as $\mathbf{g}$, in which elements $g_{f_l(j)}=g_j^{(l)}$ for $j=1,\ldots,G_l$ and the rest of the $M-G_l$ elements are zero.

Decoding of GNC codes is performed by solving linear systems of equations. We refer to a received packet whose EV is not in the span of EVs of the previously received packets of the destination node as an \textit{innovative packet} and the EV is referred to as an \textit{innovative EV}. The receiver has to receive $M$ innovative EVs to recover all the source packets.

Let $N'$ be the number of randomly encoded packets that contain $M$ linearly independent EVs among them. We define $\varepsilon_c=(N'-M)/M$ as the \textit{code overhead}. Let $N''$ denote the number of received packets among which $M$ innovative packets can be obtained in a transmission session. We refer to $\varepsilon_{cn}=(N''-M)/M$ as the \textit{code-and-network-induced overhead}. This overhead may be caused by either the encoding at the source node or scheduling and re-encoding at intermediate nodes, or both. Supposing that the decoding succeeds after receiving $N\geq N''$ packets, $\varepsilon_d=(N-N'')/M$ is referred to as the \textit{decoder-induced overhead}. Zero decoder-induced overhead is achieved if the decoder succeeds immediately once $M$ innovative packets are received. The overall reception overhead of the session is $\varepsilon=(N-M)/M=\varepsilon_{cn}+\varepsilon_d$. Note that transmission losses (packet erasures) are not counted in the overheads defined so far.

The \textit{decoding cost} of GNC codes is measured as the number of finite field arithmetic operations required for successful decoding, where an \textit{operation} refers to either a divide or a multiply-and-add between two elements of a finite field.

\section{Decoding Algorithms of GNC Codes}\label{decoder_desp}
We first introduce the G-by-G decoder and identify its decoder-induced overhead. A baseline approach and the OA decoder are then proposed to improve efficiency.
\subsection{G-by-G Decoder}
\begin{definition*}[G-by-G Decoding]
At the beginning of each \textit{step}, the decoder attempts to select a generation $\mathcal{G}_l, 1\leq l\leq L$ and decode it by solving a system of linear equations using Gaussian elimination. Denote successive rows of GEVs and information symbols of received coded packets belonging to $\mathcal{G}_l$, as $\mathbf{A}_l$ and $\mathbf{B}_l$, respectively. Decoding solves $\mathbf{A}_l\mathbf{X}_l=\mathbf{B}_l$, where rows of $\mathbf{X}_l$ are the to-be-decoded source packets in $\mathcal{G}_l$. A generation $\mathcal{G}_l$ whose $\mathbf{A}_l$ is full-rank is called \textit{separately decodable}. When a generation is decoded, decoded source packets are subtracted from the received packets of remaining undecoded generations which contain the decoded packets. This marks the end of one decoding step. Since the subtraction may reduce the numbers of unknown packets of other generations, new separately decodable generations may be found. If this is the case, the next decoding step can begin; if no such generations can be found, the decoder waits until more packets are received such that a new separately decodable generation is found. Decoding proceeds as above until all generations are decoded.
\end{definition*}

The following example illustrates that G-by-G decoding has decoder-induced overhead.
\begin{example}[Inefficiency of G-by-G decoding]\label{nonzero_decoder_overhead_example}
Assume that $4$ source packets are grouped into two generations $\mathcal{G}_1=\{\mathbf{s}_1,\mathbf{s}_2, \mathbf{s}_3\}$ and $\mathcal{G}_2=\{\mathbf{s}_2,\mathbf{s}_3, \mathbf{s}_4\}$. Suppose that $2$ packets have been received for each generation, written as $\mathbf{r}_1^{(1)}=\mathbf{s}_1+\mathbf{s}_2$, $\mathbf{r}_2^{(1)}=\mathbf{s}_2+\mathbf{s}_3$ and $\mathbf{r}_1^{(2)}=\mathbf{s}_2+\mathbf{s}_4$, $\mathbf{r}_2^{(2)}=\mathbf{s}_2+\mathbf{s}_3+\mathbf{s}_4$, respectively. In this case, $\mathbf{r}_1^{(1)}$, $\mathbf{r}_2^{(1)}$, $\mathbf{r}_1^{(2)}$ and $\mathbf{r}_2^{(2)}$ are linearly independent. However, neither generation is separately decodable and the G-by-G decoding process cannot start.
\end{example}

\subsection{Straightforward Overhead-Optimized Decoding}
The inefficiency of G-by-G decoding arises from separate decoding of each generation and that possible overlaps among generations are not exploited until a generation has been separately decoded. To resolve the issue, a straightforward overhead-optimized approach may be used by solving $\mathbf{A}\mathbf{X}=\mathbf{B}$ using GE where successive rows of $\mathbf{A}$ and $\mathbf{B}$ are the EVs and information symbols of the received coded packets, respectively, and $\mathbf{X}$ contains all the $M$ source packets. In this case, decoding is successful as soon as $M$ innovative packets are received, resulting in zero decoder-induced overhead. We refer to this as the \textit{naive decoder}. Unfortunately, the naive decoder does not take EV sparsity into account, and therefore the decoding cost may be high.

\subsection{Overlap-Aware Decoder}
We now propose an \textit{overlap-aware} (OA) overhead-optimized decoder for GNC codes with the same overhead as the naive decoder but at a lower computational cost. By being overlap aware, the OA decoder is better able to exploit overlaps among generations. 

OA decoding is first performed \textit{locally} in each generation. Let $\mathbf{A}_l$ denote the decoding matrix of $\mathcal{G}_l,l=1,\ldots, L$ which is initialized as a $G_l\times G_l$ zero matrix; $\mathbf{A}_l$ is referred to as the \textit{local decoding matrix} (LDM) of $\mathcal{G}_l$. Forward row operations are performed on the successively received GEVs. Each newly received GEV, if it is not zero after being processed against previously received GEVs, is referred to as an \textit{innovative GEV} of $\mathcal{G}_l$. The vector is stored as the $i$-th row of $\mathbf{A}_l$ if its $i$-th element is the left-most nonzero. Note that an innovative GEV may not be an innovative EV. We declare the decoder as \textit{OA ready} when a total of $M$ innovative GEVs have been received.

When OA ready, the decoder attempts to \textit{jointly} decode generations. Each $\mathbf{A}_l,l=1,\ldots,L$ is \textit{partially diagonalized} by eliminating elements above nonzero diagonal elements, rendering $\mathbf{A}_l$ sparser. The $M$ innovative GEVs are then converted to EVs to populate an $M\times M$ \textit{global decoding matrix} (GDM) $\mathbf{A}$. 

Two properties of $\mathbf{A}$ are noted. First, in each length-$M$ row of $\mathbf{A}$ there are at most $\max\{G_l,\forall l\}$ nonzero elements. Since $\max\{G_l\}\ll M$, $\mathbf{A}$ tends to be sparse. Second, if any generation is already separately decodable, its LDM has been fully diagonalized, resulting in some \textit{singleton} rows in $\mathbf{A}$, i.e., rows containing only one nonzero element. The joint decoding aims to transform $\mathbf{A}$ to an identity matrix. Exploiting its sparseness, we pivot $\mathbf{A}$ to minimize the computational cost.

\subsection{Pivoting $\mathbf{A}$ in OA Decoding}\label{solve_sparse}
Pivoting reorders rows and columns of a sparse matrix such that the computational cost of row reductions can be reduced. Finding the globally optimal pivoting sequence that minimizes the computational cost is known to be an \textit{NP-complete} problem \cite{Rose1978}. Here we only consider local heuristic methods. We propose the following method that employs two rounds of pivoting:

\subsubsection{First Round}
\begin{figure}[htbp]
\centerline{\includegraphics[width=2.5in]{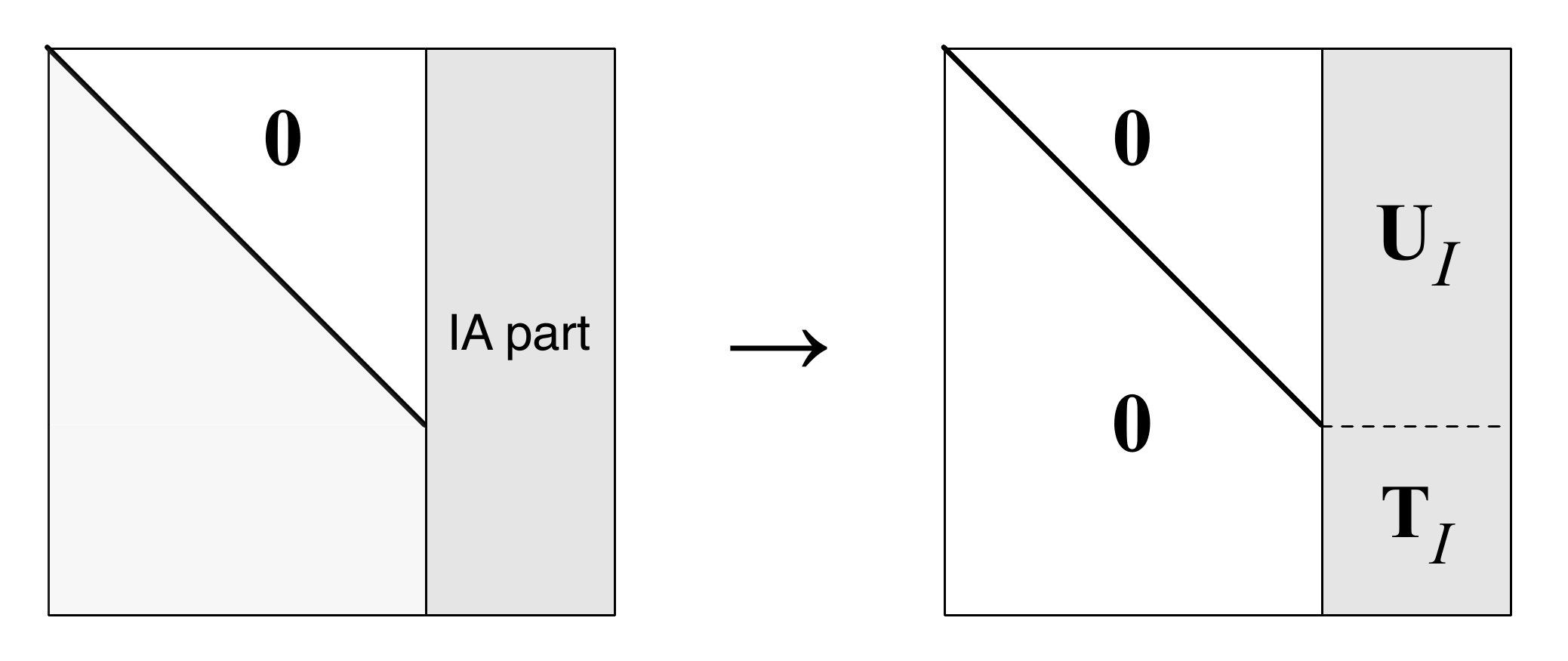}}
\caption{\label{inactivation}Reordering and partial diagonalization of $\mathbf{A}$ using inactivation pivoting.}
\end{figure}

We pivot $\mathbf{A}$ using an \textit{inactivation} approach \cite{Odlyzko1985}, \cite{Pomerance1992}, as is employed in decoding standardized raptor codes \cite{ShokrollahiLuby09}. The reordered matrix after pivoting consists of a lower triangular (\textit{active part}) and some other dense ``inactive'' columns (\textit{inactive part}), as shown in the left of Fig. \ref{inactivation} where $\mathbf{0}$ refers to areas consisting of only zero elements. The time complexity of inactivation pivoting is $\mathcal{O}(n)$ for an $n\times n$ matrix. The lower triangular part will then be diagonalized, and we refer to the sub-matrices of the inactive part of $\mathbf{A}$ as $\mathbf{U}_I$ and $\mathbf{T}_I$, respectively. The number of inactivated columns is denoted as $M_I$.

\subsubsection{Second Round}
The $\mathbf{T}_I$ of Fig. \ref{inactivation} becomes dense due to row reductions to diagonalize the active part of $\mathbf{A}$. However, the structure of GNC codes is such that the nonzero elements in the inactive part might not be uniformly located, and $\mathbf{T}_I$ may still be sparse. We perform another round of pivoting on $\mathbf{T}_I$ to exploit its sparsity.

The second round uses a modified \textit{Markowitz criterion} due to Zlatev \cite{Zlatev1980}. The original Markowitz criterion \cite{Duff1986} selects a nonzero $\mathbf{T}_I[i][j]$ as the pivot if it has the smallest \textit{Markowitz count} of the matrix, defined as $(r_i-1)(c_j-1)$, where $r_i$ and $c_j$ are the number of nonzero elements on the corresponding row and column, respectively. Instead of searching $n-i+1$ rows for the $i$-th pivot, Zlatev pivoting searches only from a constant number ($\leq 3$) of rows with the least number of nonzero entries. Therefore, the time complexity of Zlatev pivoting is also $O(n)$ for an $n\times n$ matrix while that of the original Markowitz criterion is $\mathcal{O}(n^2)$.  

After the second round of pivoting, we first perform forward row operations on the reordered $M_I\times M_I$ matrix $\mathbf{T}_I$. If the $M$ innovative GEVs that populated GDM $\mathbf{A}$ are all innovative EVs, $\mathbf{T}_I$ is full-rank. The resulting upper triangular $\mathbf{T}_I$ can then be reduced to an identity matrix. This solves $M_I$ source packets. Further eliminating $\mathbf{U}_I$ by subtracting the $M_I$ decoded packets will recover the rest of the source packets. If some innovative GEVs are not innovative EVs, the $\mathbf{T}_I$ after row operations would be an upper triangular matrix containing zero diagonal elements in some rows. In this case, more coded packets need to be received and processed to fill in the rows before reducing $\mathbf{T}_I$ to an identity matrix. 

\subsection{OA Decoder-Induced Overhead}\label{oaoh}
Only one or zero innovative EV may be generated from an innovative GEV. Suppose that $M$ innovative EVs are received, the decoder must be OA ready. With $M$ innovative EVs, the constructed $\mathbf{A}$ is full-rank. Since pivoting does not change the matrix rank, OA decoding is guaranteed to succeed and therefore has zero decoder-induced overhead.

\section{Code Design and OA Decoding}\label{design}
\subsection{Code Description}\label{code_description}
We present a code design based on the \textit{random annex code} (RAC), originally proposed in \cite{Li2011a} for the G-by-G decoder. The design of code parameters for the OA decoder, however, differs significantly from that for the G-by-G decoder. 

\begin{definition*}[Random Annex Code \cite{Li2011a}]
The $M$ source packets, $\mathcal{S}=\{\mathbf{s}_1,\ldots,\mathbf{s}_M\}$, are first partitioned into $L$ disjoint subsets $\mathcal{B}_l=\{\mathbf{s}_{(l-1)B+1},\ldots,\mathbf{s}_{lB}\},l=1,\ldots,L$ of equal size $B$ (we assume $M=LB$, i.e., $L$ to be a divisor of $M$; otherwise null packets can be used for padding), one per generation. $\mathcal{B}_l$ is referred to as the \textit{base part} of the generation. After that, each generation $\mathcal{G}_l$ is equipped with a random \textit{annex} of size $H$, denoted as $\mathcal{H}_l$, which consists of a random selection of $H$ packets from $\mathcal{S}-\mathcal{B}_l$. The annex code introduces overlap between generations. The overall generation $\mathcal{G}_l=\mathcal{B}_l\cup\mathcal{H}_l$ and $G=|\mathcal{G}_l|=B+H$. When generating a coded packet, one generation is chosen uniformly at random.
\end{definition*}

The proposed design has two customizations relative to the original RAC: 1) \textit{precoded} source packets using a binary systematic erasure-correction code and 2) only the binary field, $\mathbb{F}_2$ is used when coding a packet. The proposed design is referred to as the \textit{precoded binary RAC} (PB-RAC).
\begin{definition*}[Precoded Binary Random Annex Code]
The $M$ source packets are first precoded using a binary systematic code to obtain $M+S$ \textit{intermediate packets}, $\{\mathbf{s}_1, \ldots, \mathbf{s}_M, \mathbf{c}_1, \ldots, \mathbf{c}_S\}$, where $\mathbf{c}_i=\sum_{j=1}^Mw_{i,j}\mathbf{s}_j, i=1,\ldots,S$ are parity-check packets, the $w_{i,j}$ are chosen from $\mathbb{F}_2$ and $S=\theta M$. We assume that each source packet is covered by at least one parity-check packet. RAC is then applied to the intermediate packets and a total of $L'=\frac{M+S}{B}$ generations are formed from the intermediate packets with base part size $B$ and generation size $G$.
\end{definition*}

\subsection{OA Decoding of PB-RAC}
The addition of a precode is inspired by the improvements obtained by adding a precode to the Luby transform (LT) code to form raptor codes \cite{Shokrollahi2006}. As with raptor codes, we assume that the PB-RAC precoding coefficients are known to all the receivers of the transmission. The precode codes across the source packets of all generations before RAC coding is performed. This helps to alleviate the coupon collector problem characterized in \cite{Li2011a}. With a systematic precode, we show below that the OA decoder efficiently affects joint decoding of the precode and RAC.

Unlike successively-decoded network codes and precodes in previous works (e.g., in \cite{Tang2012}, the decoder needs to recover a pre-defined fraction of intermediate packets before decoding of the precode can begin), OA joint decoding begins the pivoting by appending the parity-check matrix of the precode to the GDM of the network code and performing pivoting on the combined matrix. Given a valid precode, joint decoding guarantees successful decoding when $M$ innovative EVs are received.

Since RAC is applied to $M+S$ intermediate packets in PB-RAC, $\mathbf{A}$ has $M+S$ columns rather than $M$ as in the non-precoded case. The $S$ parity-check constraint equations of the precode, i.e., $\sum_{j=1}^M w_{i,j}\mathbf{s}_j+\mathbf{c}_i=0$ for $i=1,2,\ldots, S$, are then appended to $\mathbf{A}$: Let $\mathbf{H}=\left[\begin{array}{cc}\mathbf{W} & \mathbf{I}_{S\times S}\end{array}\right]$, where elements $w_{i,j}$ of $\mathbf{W}$, $1\leq i\leq S$, $1\leq j\leq M$ are coding coefficients of the systematic precode and $\mathbf{I}_{S\times S}$ is the size-$S$ identity matrix. We therefore have
\begin{equation}
\left[\begin{array}{c}\mathbf{A}\\ \mathbf{H}\end{array}\right]\left[\begin{array}{cccccc}\mathbf{s}_1^T & \cdots & \mathbf{s}_M^T & \mathbf{c}_1^T & \cdots & \mathbf{c}_S^T\end{array}\right]^T=\left[\begin{array}{c}\mathbf{B}\\ \mathbf{0}_{S\times K}\end{array}\right],
\end{equation}
where $[\cdot]^T$ denotes transpose and $\mathbf{0}_{S\times K}$ is the $S\times K$ zero matrix. Pivoting is performed on the binary $(M+S)\times(M+S)$ matrix $\mathbf{A}_{\text{eff}}=\left[\begin{array}{cc}\mathbf{A}^T & \mathbf{H}^T\end{array}\right]^T$. If the $M$ innovative GEVs are all innovative EVs, a full-rank $\mathbf{A}_{\text{eff}}$ can be obtained with high probability when a valid precode is used.

\subsection{Choosing Parameters for PB-RAC}\label{PB-RAC_analysis}
In this subsection we choose code parameters for PB-RAC. First, we analytically show that precoding (i.e., $S>0$) and/or allowing for overlap (i.e., generation size $G>B$) help reduce code overhead, which is inversely proportional to the probability that innovative coded packets are received.

Let $p_k$ be the probability that the next received coded packet is innovative for the receiver when $k<M$ coded packets have been received. Let $\mathrm{Pr}(n_1,\ldots,n_{L'};k)$ denote the probability of any combination of $n_1,\ldots,n_{L'}$ received packets satisfying $\sum_{l=1}^{L'}n_l=k$ where $n_l\geq 0$ is the number of received packets belonging to the $l$-th generation among the $L'=\frac{M+S}{L/M}=\frac{M+S}{B}$ generations. We assume that packets belonging to each generation arrive at the receiver with equal probability $\frac{1}{L'}$, which reflects that the generations are scheduled with equal likelihood. The above joint probability follows the multinomial distribution,
\begin{eqnarray}
  \mathrm{Pr}(n_1,\ldots,n_{L'};k) &\triangleq& \mathrm{Pr}(\mathbf{n};k)\nonumber\\
                         &=& \frac{k!}{n_1!\cdots n_{L'}!}\left(\frac{1}{L'}\right)^{n_1}\cdots \left(\frac{1}{L'}\right)^{n_{L'}}\nonumber\\
                         &=& \frac{k!}{n_1!\cdots n_{L'}!}\left(\frac{1}{L'}\right)^{k}.
\end{eqnarray}

Considering the precoded but non-overlapping case first, i.e., $S>0, G=B$, to simplify the analysis, we assume that a sufficiently large finite field is used. Note that innovative GEVs are equivalent to innovative EVs in the non-overlapping case. When $k=\sum_{l=1}^{L'}n_l$ packets have been received, the next packet would be non-innovative only if it belongs to a generation that has at least $B$ received packets. Let $\mathcal{S}$ be the set of all $\mathbf{n}$'s with $\sum_{l=1}^{L'}n_l=k$ and $u(\mathbf{n};B)$ be the number of elements in $\mathbf{n}$ that are greater than or equal to $B$. For a given $\mathbf{n}\in\mathcal{S}$, $u(\mathbf{n};B)/L'$ is the probability of the next received packet being non-innovative. Hence,
\begin{eqnarray}\label{pk_precode}
p_k &=& 1-\frac{\sum_{\mathbf{n}\in\mathcal{S}}\mathrm{Pr}(\mathbf{n};k)u(\mathbf{n};B)}{L'}\nonumber\\
    &=& 1-\frac{\displaystyle\sum_{\mathbf{n}\in\mathcal{S}} \frac{k!}{n_1!\cdots n_{L'}!}\left(\frac{1}{L'}\right)^{k} u(\mathbf{n};B) }{L'}.
\end{eqnarray}
Note that (\ref{pk_precode}) also applies to the non-precoded case, replacing $L'$ with $L$. Given that $L'>L$, $p_k$ of the precoded case is therefore greater than that of the non-precoded case, i.e., code overhead is reduced after precoding.

When we allow for overlap, i.e., $S>0, G>B$, we are not able to obtain an exact expression for $p_k$ (because an innovative GEV does not guarantee an innovative EV). However, we can lower bound it using (\ref{pk_precode}) because packets are encoded across more intermediate packets in the overlapping case, and therefore its $p_k$ is always greater than that of the precoded but non-overlapping case, i.e.,
\begin{equation}\label{pk_overlap_lower}
  p_k>1-\frac{\displaystyle\sum_{\mathbf{n}\in\mathcal{S}} \frac{k!}{n_1!\cdots n_{L'}!}\left(\frac{1}{L'}\right)^{k} u(\mathbf{n};B) }{L'}.
\end{equation}
Similar to (\ref{pk_precode}), (\ref{pk_overlap_lower}) applies to $S=0$ as well. Therefore, allowing for overlap is capable of reducing code overhead in both the non-precoded and precoded cases. We note that $p_k$ for allowing overlap is upper bounded by
\begin{equation}\label{pk_overlap_upper}
  p_k< 1-\frac{\displaystyle\sum_{\mathbf{n}\in\mathcal{S}} \frac{k!}{n_1!\cdots n_{L'}!}\left(\frac{1}{L'}\right)^{k} u(\mathbf{n};G) }{L'},
\end{equation}
where the right-hand side corresponds to the probability assuming that the first $G$ packets received by each generation are always innovative.

\begin{figure}[htbp]
\centerline{\includegraphics[width=4in]{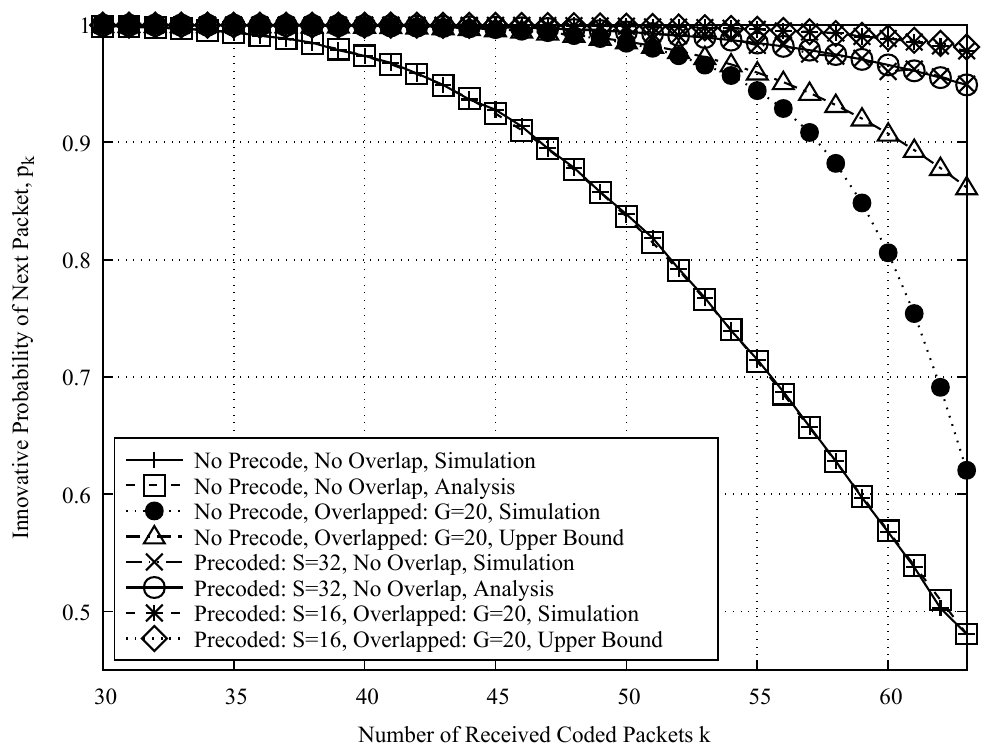}}
\caption{\label{pk_comparisons}Comparisons of $p_k$ for $k\geq 30$ of various codes; $M=64$ and $B=16$.}
\end{figure}

A comparison of $p_k$'s for the different cases is shown in Fig. \ref{pk_comparisons} with (\ref{pk_precode}), (\ref{pk_overlap_upper}), and Monte Carlo simulation results plotted. The simulation uses $\mathbb{F}_{256}$ and each parity-check packet of the precode is a random linear combination of all the source packets. The simulations of the non-overlapping cases match the analysis accurately and that of using $S=16, G=20$ approaches the upper bound of (\ref{pk_overlap_upper}). Compared to the no-precode-no-overlap case, Fig. \ref{pk_comparisons} shows that increasing $S$ and/or $G$ results in higher $p_k$ and hence reduces code overhead\footnote{It is noted that $G$ cannot be increased without bound when designing RAC for the G-by-G decoder \cite{Li2011a}, \cite{liye2013}.}. However, we argue that increasing both $S$ and $G$ somewhat is preferable to increasing only one of them and can achieve a similar $p_k$. For example as shown in Fig. \ref{pk_comparisons}, using $S=16, G=20$ results in similar (even slightly higher) $p_k$ than using $S=32$ but no overlap. A mixed use better balances the matrix sparseness of $\mathbf{A}_{\text{eff}}=\left[\begin{array}{cc}\mathbf{A}^T & \mathbf{H}^T\end{array}\right]^T$, leading to fewer columns to inactivate and hence lower decoding cost as shown in Section \ref{section: evaluation}.

Our choices of $S$ and $G$ are as follows. For a given $M$, we use the same systematic LDPC precode as standard raptor codes \cite{Luby2007} (Section 5.4.2.3), which has a fixed efficient structure and is suitable for inactivation pivoting. The value of $S$ therein is the smallest prime number greater than or equal to $\lceil0.01M\rceil+X$ where $X$ is the smallest integer such that $X(X-1)\geq 2M$. To determine $G$, we first fix $B=32$ as it is demonstrated in \cite{Paramanathan2013} that generations of size $32$ yield acceptable GE decoding speed when performing RLNC over $\mathbb{F}_2$. Given $B$ and $S$, noting that $p_k$ with $G>B$ may approach the upper bound (\ref{pk_overlap_upper}) as shown in Fig. \ref{pk_comparisons}, $G$ is therefore chosen to satisfy the following criteria: when a total of $M$ packets are received where each packet may belong to any generation with equal probability, the probability of a generation receiving more than $G$ packets is sufficiently small (see below). As discussed above, the numbers of packets received by the generations follow a multinomial distribution. To simplify calculation of the probability, as an approximation, we may instead model the numbers of received packets of the generations as IID Poisson random variables with rate parameter $\tau=M/L'$ \cite{Mitzenmacher:2005:PCR:1076315}. Let $Y$ denote the number of received packets of a generation. We propose to find the smallest $G$ (in favor of sparseness) satisfying the criterion expressed as:
\begin{equation}
G_{\text{min}}=\inf \left\{G\in\mathbb{Z}^{+}: \mathrm{Pr}\{Y>G\}<\frac{1}{L'}\right\},
\end{equation}
where the inequality ensures that the expected number of generations that receives more than $G_{\text{min}}$ packets is less than one. $G_{\text{min}}$ can be found by an integer search starting from $G=B$.

\subsection{Number of Inactivated Columns} 
From Fig. \ref{inactivation}, we see that the cost of OA decoding depends heavily on the number of inactivated columns $M_I$ because the $M_I\times M_I$ matrix $T_I$ needs to be reduced to an identity matrix using GE. In this subsection, we present a method to estimate the fraction of columns that would be inactivated when pivoting $\mathbf{A}_{\text{eff}}$ for a given set of parameters $\{M, S, B, G\}$. For ease of analysis, we make the following further assumptions: 1) a set of parameters $\{M, S, B, G\}$ is chosen so that the first $M$ received packets are linearly independent; 2) after uniformly randomly inactivating $\alpha M$ columns in $\mathbf{A}$, $\alpha\in(0,1)$, G-by-G decoding can successfully decode $(1-\delta)L'$ generations (or equivalently $M$ intermediate packets) with $M$ received packets, where $\delta\equiv\frac{\theta}{1+\theta}$, $\theta=S/M$.

The above assumptions ensure that $\mathbf{A}_{\text{eff}}$ can be pivoted successfully with a total of $M_I=(\alpha+\theta)M$ columns being inactivated (i.e., inactivating another $S=\theta M$ columns after appending the parity-check rows $\mathbf{H}$ to $\mathbf{A}$). Note that the rest of OA decoding only involves back substitutions on $\mathbf{A}_{\text{eff}}$ after the inactivated part is solved via GE. The whole OA decoding process can therefore be viewed as a combination of inactivation and G-by-G decoding. 

The expected proportion of inactivated columns, $\alpha$, can be determined using an asymptotic analysis of G-by-G RAC decoding. In order to recover $(1-\delta)L'$ generations using the G-by-G decoder when no inactivation is introduced, the following inequality has to be satisfied for all $x\in[\delta,1]$:
\begin{equation}\label{iterative_governer}
\sum_{u=0}^{G-1}\eta_{u}\sum_{k=u}^{G-1}\left(\begin{array}{c}G-1\\ k\end{array}\right)(\lambda(x))^k(1-\lambda(x))^{G-1-k}<x,
\end{equation}
where $\eta_{u}$ is the probability that one generation has $u$ innovative packets received. The detailed derivation of (\ref{iterative_governer}) is provided in Appendix A where also $\lambda(x)$ is defined in (\ref{lambda_asym}).

Now if we inactivate $\alpha M$ packets in $L'$ generations, on average $\alpha M/L'$ packets per generation, then the G-by-G decoding only needs to ``decode'' $G'=G-\alpha M/L'$ packets from each generation (the ``decoded'' packets will be linear combinations of the inactivated packets of the generation). Therefore, we need to modify (\ref{iterative_governer}) by replacing $G$ with $G'$. The inequality needs to be satisfied for all $x\in[\delta,1-\alpha M/(M+S)]$.

Again, we model the number of received packets of each generation as independent and identically distributed (IID) Poisson variables with rate parameter $\tau=M/L'$. Therefore $\eta_u=\frac{\tau^u\mathrm{e}^{-\tau}}{u!}$ in (\ref{iterative_governer}). Let $f(M, S, B, G, \alpha)$ denote the modified left-hand side of (\ref{iterative_governer}) after replacing $G$ with $G'$. The required value of $\alpha$ is
\begin{equation}\label{ia_fraction_analysis}
\alpha^{\ast}=\inf\left\{\alpha:f(M, S, B, G, \alpha)<x,\ x\in[\delta,1-\alpha M/(M+S)]\right\}.
\end{equation}
We find $\alpha^{\ast}$ using one-dimensional exhaustive search starting from $\alpha=0$ with desired precision increment $\Delta\alpha$. In each search step, the inequality in (\ref{ia_fraction_analysis}) is tested by discretizing $x$ in $[\delta,1-\alpha M/(M+S)]$. The total number of inactivated columns is then $M_I=(\alpha^{\ast}+\theta)M$.
 
\section{Performance Evaluation}\label{section: evaluation}
We now evaluate the performance of our design. Throughout we use packets each containing $K=1600$ one-byte source symbols. Coding coefficients used for precoding and RLNC at the source or intermediate nodes are from $\mathbb{F}_2$ unless stated otherwise. We count the total number of operations on both sides of the linear system of equations in our developed software library \cite{libslnc}, denoted as $N_{\text{ops}}$. We use the \textit{average number of operations per symbol}, $\frac{N_{\text{ops}}}{MK}$, to compare decoding costs. Performance results are reported below based on averaging over $1000$ trials.

\subsection{Code and Decoder Performance}
\begin{figure}[htbp]
\centerline{\includegraphics[width=4in]{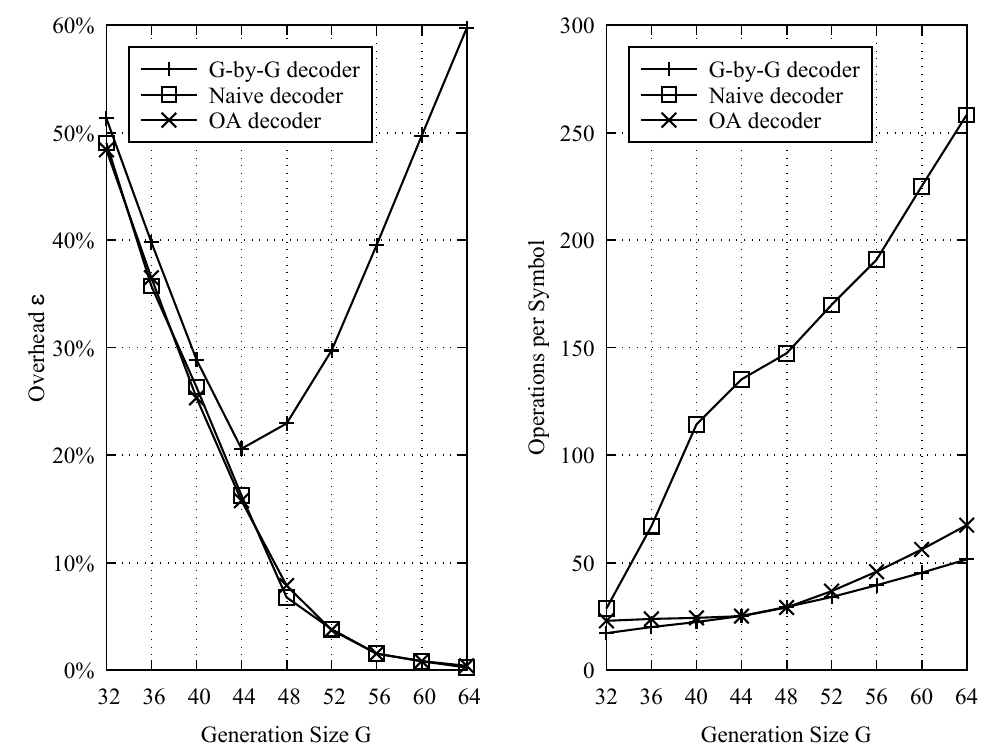}}
\caption{\label{Iterative_Naive_OA_decoders}Comparison of G-by-G, naive and OA decoders for RAC with $M=1024$, $B=32$ and $L=32$.}
\end{figure}

We first evaluate the design in the scenario where the decoders operate on GNC-coded packets directly and therefore only code and decoder-induced overheads may be incurred. In Fig. \ref{Iterative_Naive_OA_decoders}, we compare the proposed OA decoder with the G-by-G and the naive decoders. We use non-precoded RAC for comparison. We show the overheads and decoding costs of RAC for different generation sizes. It is seen that the OA decoder has the same overhead as the naive decoder. For G-by-G decoding, the introduction of overlap initially alleviates the ``coupon collector problem" \cite{Li2011a} to efficiently reduce the overhead to $20\%$. Beyond this, the inherent limitation of G-by-G strategy increases the overhead. The overheads of the naive and OA decoders are due only to code overhead. The decoding cost of G-by-G decoder is the lowest and increases only slightly with generation size. The OA decoding cost is much lower than that of the naive decoder and is close to that of the G-by-G decoder.

\begin{figure}[htbp]
\centerline{\includegraphics[width=4in]{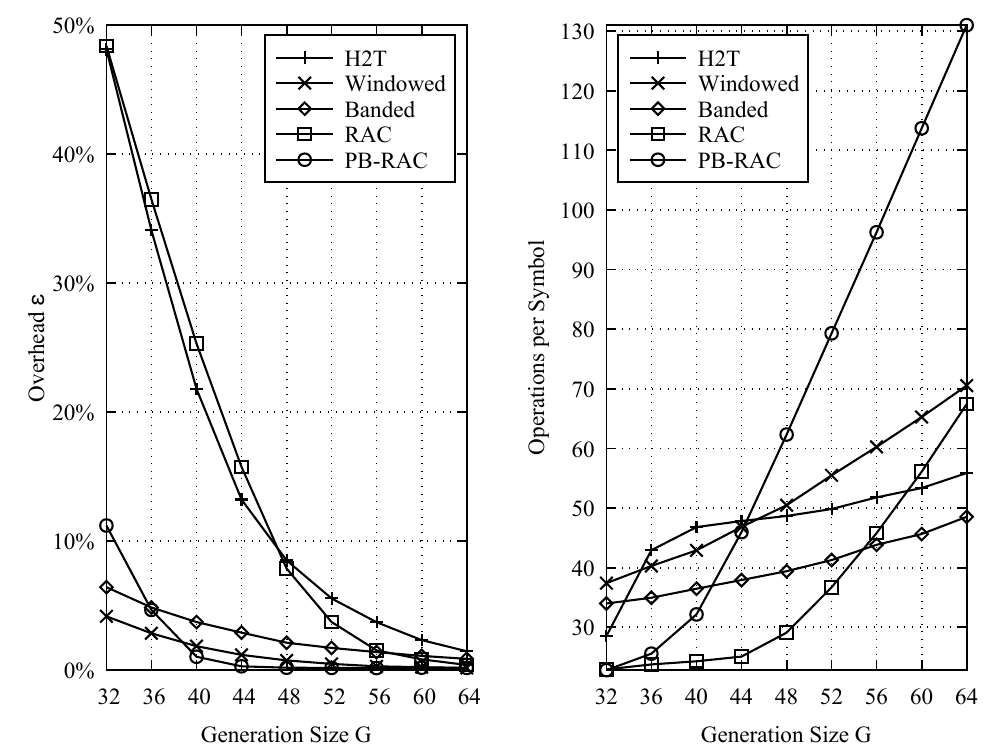}}
\caption{\label{H2T_RAC_PB-RAC}Comparison of H2T from \cite{Heidarzadeh2010}, \cite{Heidarzadeh2011}, windowed codes from \cite{Heide2014}, banded codes from \cite{Fiandrotti2014}, RAC, and PB-RAC; $M=1024$ and $B=32$, $S=59$ for PB-RAC.}
\end{figure}

Next we compare the proposed PB-RAC with the original RAC of \cite{Li2011a}, \textit{head-to-toe} (H2T) codes of \cite{Heidarzadeh2010}, \cite{Heidarzadeh2011}, \textit{windowed} codes of \cite{Heide2014}, and \textit{banded} codes of \cite{Fiandrotti2014}. The H2T codes have the same number of $L=M/B$ generations as RAC. The generations of H2T codes are overlapped consecutively rather than randomly. A windowed code can be viewed as a H2T code with $M$ generations. The banded code is similar to the windowed code except that it does not allow for wrap-around and therefore has only $M-G+1$ generations. The decoding matrices of banded codes are strictly banded while those of H2T and windowed codes are close to banded (except for the last few rows which wrap around). In Fig. \ref{H2T_RAC_PB-RAC}, we show the overheads and decoding costs of the four codes for different generation sizes. Note that the probability of sending coded packets from each generation of the banded code is not uniform as in \cite{Fiandrotti2014} while that of the other three codes are uniform. The decoding of H2T codes, windowed codes and banded codes utilize straightforward GE whereas the decoding of RAC and PB-RAC use the proposed OA decoder.

As generation size $G$ grows, overheads of all codes decrease. However, PB-RAC has much lower overhead than that of H2T and RAC, achieving less than $1\%$ at $G=41$ while RAC achieves this at $G=58$ and H2T at $G>64$. It is important to note that overheads of H2T decrease more slowly when approaching zero than that of RAC, which is a major justification for our choice of RAC for design. The decoding costs of achieving $1\%$ overheads for PB-RAC, RAC and H2T are $35$, $50$, and $60$ operations per symbol, respectively. A significant improvement in both overhead and decoding cost is obtained using PB-RAC. It is interesting to note that windowed and banded codes have very similar code overheads to that of PB-RAC even though no precoding is used. The windowed and banded code achieve $1\%$ overhead at $G=45$ and $G=61$, respectively, and both require about $47$ operations per symbol to decode. However, subsequent results show that the two codes might not be suitable for use in networks where intermediate nodes need to re-encode packets on-the-fly during transmission. 

We have noted that OA decoding costs of RAC and PB-RAC increase quickly with generation size, while those of H2T, windowed and banded codes are much flatter. This is expected since the decoding matrices of RAC and PB-RAC, being less structured, require more decoding operations to decode even though OA decoding has been used to exploit sparseness. Nevertheless, since the overheads of H2T, windowed and banded codes decrease more slowly as the generation size increases, to achieve the same low level of overhead PB-RAC with OA decoding requires fewer operations because the required code can be much sparser. 

In Table \ref{achieve_1_percent_ohs} we show decoding performances of four PB-RAC's that result in about $1\%$ overhead, where $S$ denotes the number of parity-check packets added by the precode and $S=0$ corresponds to the non-precoded case (i.e., original RAC). The trade-off between $S$ and $G$ and a suitable choice of $S$ and $G$ yielding the least number of inactivated columns and, as a consequence, the least cost, are demonstrated.
\begin{table}[h]
\begin{center}
\begin{tabular}{|c|c|c|c|c|}
\hline
$S$ & $G$ & $\varepsilon$ & $\frac{N_{\text{ops}}}{MK}$ & $M_I$\\
 \hline
0 & 58 & 0.92\% & 50 & 127\\
\hline
\textbf{59} & \textbf{41} & \textbf{0.74\%} & \textbf{35} & \textbf{80}\\
\hline
101 & 39 & 0.70\% & 38 & 100\\
\hline
149 & 37 & 0.95\% & 39 & 112\\
 \hline
\end{tabular}
\end{center}
\caption{Comparison of decoding cost $\frac{N_{\text{ops}}}{MK}$ and number of inactivated columns $M_I$ of PB-RACs achieving $\varepsilon=1\%$ with different $S$; $M=1024$, $B=32$.}\label{achieve_1_percent_ohs}
\end{table}

\begin{table}
\begin{center}
\begin{tabular}{|c|c|c|c|c|}
\hline
$M$ & 1024 & 4096 & 7168 & 10240\\
\hline
\hline
$S$ & 59  & 137 & 193 & 251\\
\hline
$G_g$ & 42 & 43 & 44 & 44\\
\hline
$G_o$ & 41 & 45 & 47 & 48\\
\hline
\end{tabular}
\end{center}
\caption{Parameters of PB-RAC codes for various numbers of source packets $M$.}\label{PB-RAC_parameters_table}
\end{table}

\begin{figure}[htbp]
\centerline{\includegraphics[width=4in]{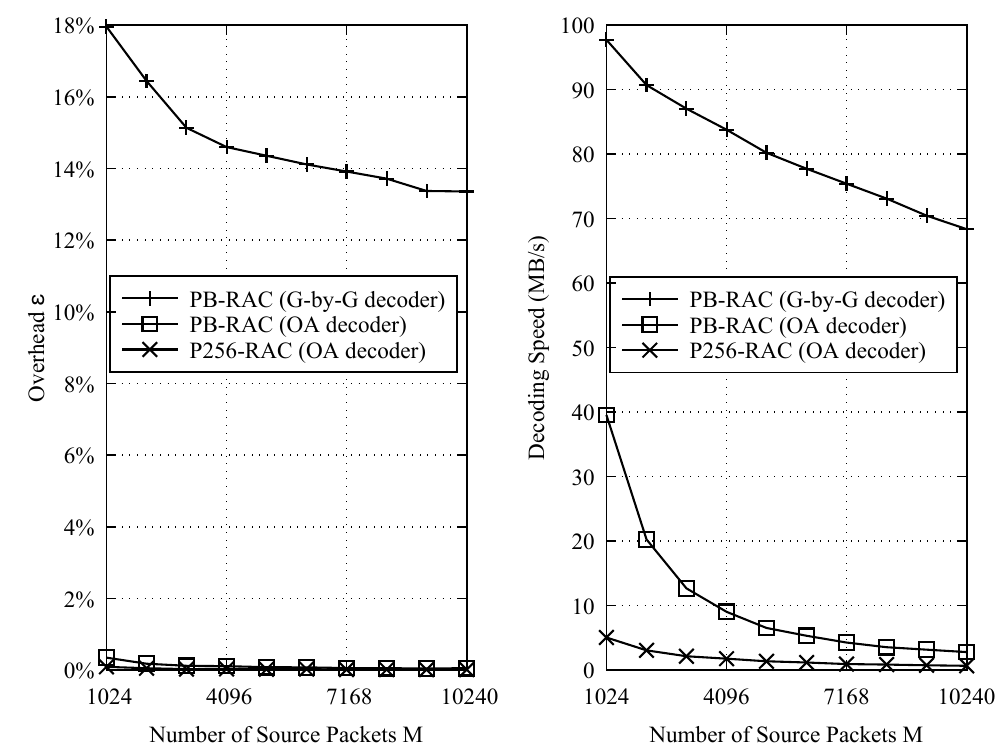}}
\caption{\label{PB-RAC_various_M}Performances of PB-RAC codes for various numbers of source packets, $M$.}
\end{figure}

We show the performances of PB-RAC for various $M$'s in Fig. \ref{PB-RAC_various_M}, where the code parameters are determined according to Section \ref{PB-RAC_analysis}. The results are compared with two other schemes: One applies G-by-G decoding on another set of PB-RAC codes with the same $B$ and $S$ but $G$ is chosen to minimize the overhead (see Fig. \ref{Iterative_Naive_OA_decoders}). The other one is P256-RAC whose coding parameters are the same as the designed PB-RAC but nonzero precode coefficients are from $\mathbb{F}_{256}-\{0\}$ and RLNC are performed in $\mathbb{F}_{256}$. We also perform OA decoding on P256-RAC. Selected coded parameters are presented in Table \ref{PB-RAC_parameters_table} where $G_g$'s and $G_o$'s are $G$ for when G-by-G and OA decoding are used, respectively. From Fig. \ref{PB-RAC_various_M}, we see that two codes achieve almost zero overhead whereas PB-RAC with G-by-G decoding has more than $12\%$ overhead. The difference in overhead by using $\mathbb{F}_2$  versus $\mathbb{F}_{256}$ is slight. We compare the decoding speed of the three schemes which is defined as the amount of data (i.e., $MK$ bytes) divided by the total CPU time needed to completely decode all the $M$ packets. The comparison (compiled with \texttt{-O3} using \texttt{gcc}) is conducted on a 2.66 GHz quad-core Intel Core 2 CPU with 4 GB RAM. The implementation is not carefully optimized and the speeds are for rough comparison. The decoding speed of G-by-G decoding is the highest and it does not decrease much as $M$ grows. The OA decoding speed is much lower and decreases rapidly as $M$ grows, suggesting that OA decoding may not be practically feasible for larger $M$; however, as $\mathbb{F}_2$ arithmetic is readily implemented in digital logic, custom hardware may be an attractive solution. The OA decoding speed of PB-RAC is considerably higher than that of P256-RAC. The OA decoding speed of PB-RAC at $M=10240$ is $3$ MB/s.

\begin{figure}[htbp]
\centerline{\includegraphics[width=4in]{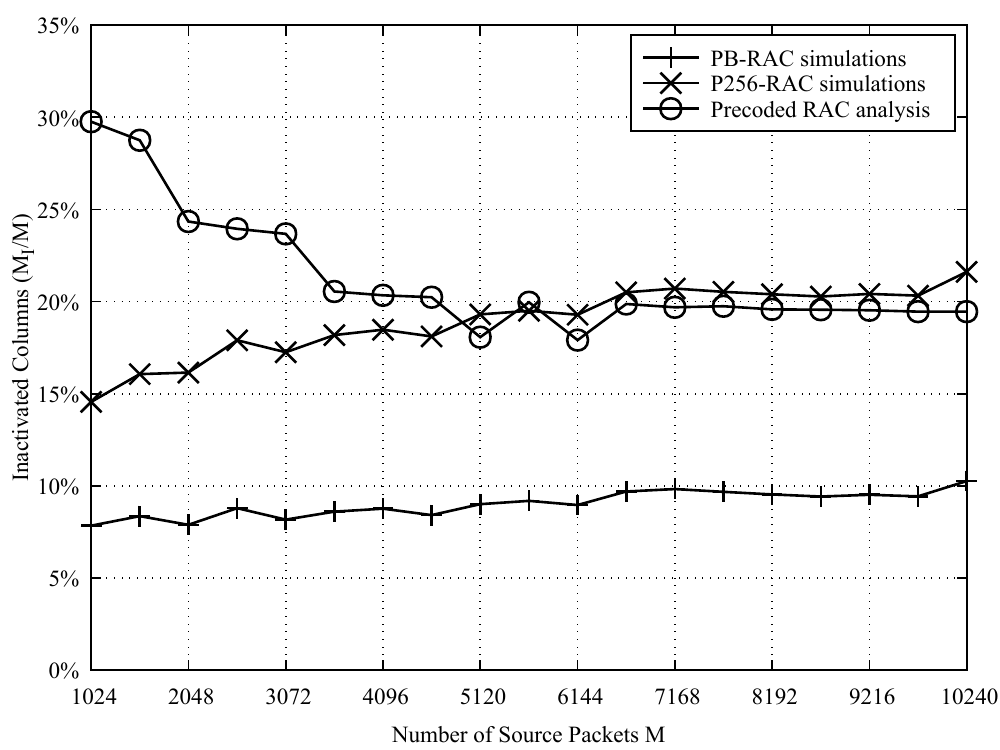}}
\caption{\label{inactivated_fractions}Inactivated columns in OA decoding of PB-RAC.}
\end{figure}

Fig. \ref{inactivated_fractions} compares the fraction of inactivated columns in OA decoding, $\frac{M_I}{M}$. The gap between P256-RAC and PB-RAC demonstrates the benefit of using $\mathbb{F}_2$: the number of inactivated columns can be reduced by half. This is because PB-RAC is much sparser with about half of its coding coefficients being zero. The analysis result ($\Delta\alpha=0.001$ when solving (\ref{ia_fraction_analysis})) which assumed sufficiently large finite field size closely matches the simulation results of P256-RAC for large $M$. For smaller $M$, it is less accurate because the analysis in Appendix A assumes sufficiently large $M$.

\subsection{Network Performance}
In this subsection we evaluate the performance of PB-RAC in the well-known lossy butterfly network \cite{Ahlswede2000}. The erasure rates of the links are set equally to $p_e=0.1$. The max-flow capacity of the network is $1.8$ packets per network use, where each \textit{network use} corresponds to both the source and intermediate nodes each sending a packet.

\begin{figure}[htbp]
\centerline{\includegraphics[width=4in]{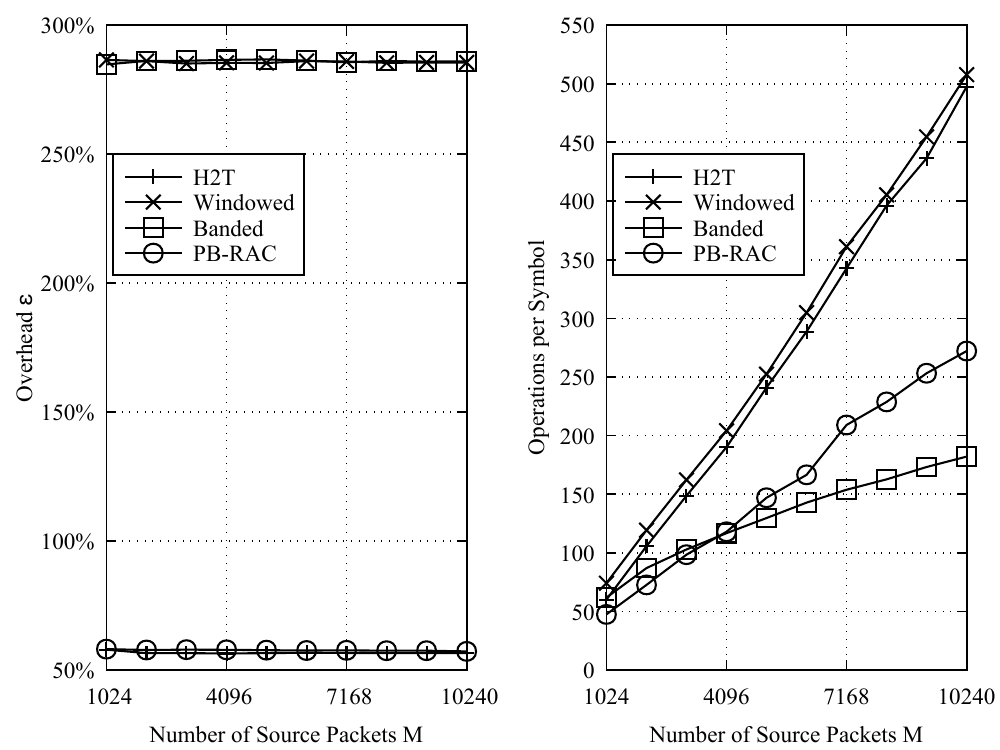}}
\caption{\label{H2T_Windowed_Banded_PB-RAC_various_M_butterfly}Comparison of H2T ($B=32$), windowed code, banded code and PB-RAC over a lossy butterfly network \cite{Ahlswede2000} where all relay nodes perform random scheduling and re-encoding in $\mathbb{F}_2$; erasure rates of links are equally $p_e=0.1$.}
\end{figure}

Fig. \ref{H2T_Windowed_Banded_PB-RAC_various_M_butterfly} shows the overheads and decoding costs when using H2T, windowed code, banded code and PB-RAC. We set $G=2\sqrt{M}$ for H2T, windowed and banded codes, as it is shown empirically in \cite{Studholme2010} that $2\sqrt{M}$ is the required width for a banded random matrix to have similar probabilistic rank properties as a dense random matrix. The parameters of PB-RAC are the same as in Fig. \ref{PB-RAC_various_M}. Each intermediate node performs random scheduling \cite{Maymounkov2006} and re-encodes packets using random coefficients from $\mathbb{F}_2$. Straightforward GE is used for decoding H2T, windowed and banded codes. PB-RAC uses OA decoding.

Fig. \ref{H2T_Windowed_Banded_PB-RAC_various_M_butterfly} shows that PB-RAC and H2T have the same overhead. All the generation sizes $G$ of PB-RAC (see Table \ref{PB-RAC_parameters_table}) are smaller than $50$ whereas those of H2T are $2\sqrt{M}\geq 64$; hence, PB-RAC is much sparser. The overheads of the windowed and banded codes, however, are high. The reason is due to limited re-encoding opportunity: the two codes each have almost $M$ generations and therefore the number of buffered packets for each generation at the relays is far fewer than that of H2T or PB-RAC, each of which have $L\ll M$ generations. We note that measures can be taken at intermediate nodes to alleviate the issue \cite{Fiandrotti2014}, \cite{Heide2014}. For example, one can search for buffered packets from different generations so that a re-encoded packet is confined to a desired window/band; or one can perform decoding at intermediate nodes to obtain a re-encoded packet. Such measures, however, complicate intermediate node processing. Since PB-RAC is much sparser, its decoding cost is lower than those of H2T and windowed codes. The decoding cost of the banded code is lowest due to its strict banded structure.

\begin{figure}[h]
\centerline{\includegraphics[width=4in]{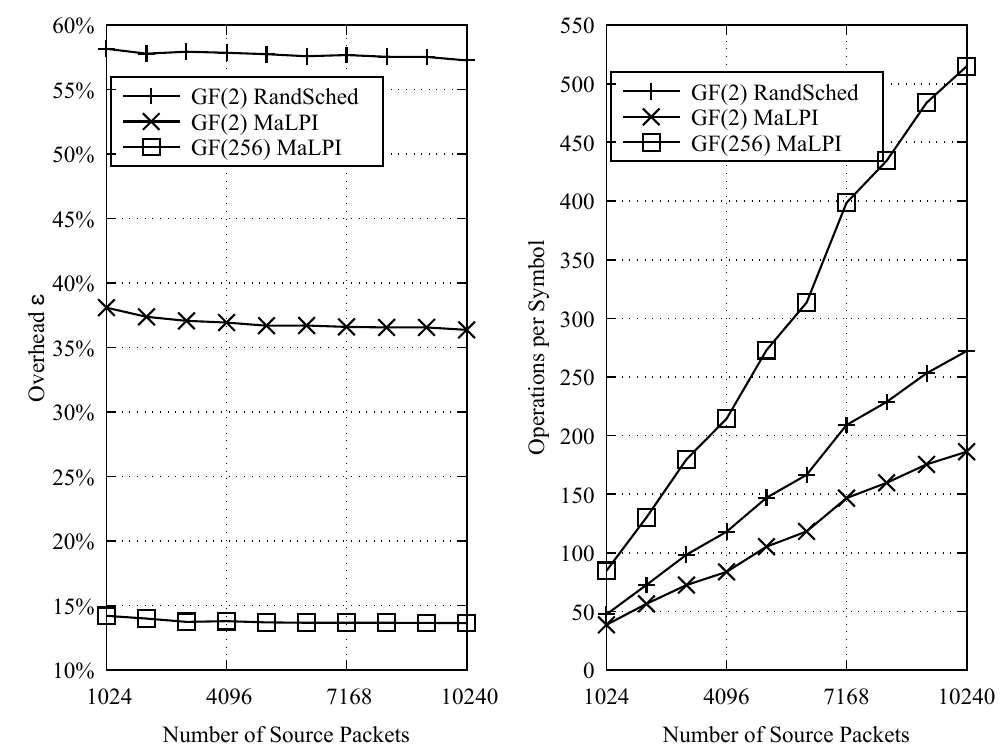}}
\caption{\label{PB-RAC_butterfly_various_schedulings}Performance of PB-RAC with different scheduling strategies at relay nodes in butterfly network; $p_{e}=0.1$ for all links.}
\end{figure}

It is seen from Fig. \ref{H2T_Windowed_Banded_PB-RAC_various_M_butterfly} that although the codes may have close-to zero code overhead and zero decoder-induced overhead (see Fig. \ref{PB-RAC_various_M}), the reception overhead over a network is nonzero because of scheduling and re-encoding at intermediate nodes. We remark that the performance may be improved by optimizing scheduling. In Fig. \ref{PB-RAC_butterfly_various_schedulings}, we show the performances of the same PB-RAC codes over the network where the MaLPI scheduling proposed in \cite{liye2013} is used at intermediate nodes. Unlike random scheduling, MaLPI chooses a generation that is \textit{least scheduled} based on the number of received packets of each generation. The overhead is reduced from $58\%$ to $38\%$. If we use $\mathbb{F}_{256}$ for re-encoding PB-RAC packets at intermediate nodes, the overheads can be further reduced to $14\%$, which correspond to the rate of about $1.57$ packets per network use (i.e., close to $90\%$ of the max-flow capacity). Note that re-encoding with $\mathbb{F}_{256}$ at intermediate nodes will change all network coding coefficients to $\mathbb{F}_{256}$, so the decoding cost will increase accordingly, as shown Fig. \ref{PB-RAC_butterfly_various_schedulings}.

\section{Conclusion}\label{summary}
In this paper we proposed an OA decoder for GNC codes. It is shown that local processing and pivoting can be employed to exploit the sparseness of GNC codes. The decoder is shown to have the same overhead as straightforward GE decoder but with much lower computational cost. For moderate numbers of source packets, the decoding cost could approach that of the linear-time G-by-G decoder. A new code that combines precoding, random overlapping generations and binary RLNC is designed and decoded by the OA decoder. It is demonstrated that a balanced combination of precoding rate and generation overlap size is crucial to obtaining low decoding costs. For a given low overhead design goal, the proposed design using PB-RAC with OA decoding is shown to have the least decoding cost compared to existing schemes.  

\appendices
\section{Analysis of G-by-G Decoding of Precoded RAC}
A precoded RAC with parameters $M$, $S$, $B$, $G$ can be viewed as a realization from an ensemble of bipartite graphs. The reader is referred to \cite{Luby1998, Richardson2001} for an exposition of the method used in this appendix. The left and right nodes of the bipartite graph correspond to intermediate packets and generations, respectively. An edge connects a pair of left and right nodes if the intermediate packet is present in the generation. The ensemble is characterized by the left-node degree distribution $\Psi(x)=\sum_{k=1}^{L'}\Psi_kx^{k}$ and the constant right-side edge degree $G$, where $L'=\frac{M+S}{B}$ and $\Psi_k$ denotes the fraction of left nodes that are of degree $k$, i.e., intermediate packets that are present in $k$ generations. According to the definition of RAC in Section \ref{code_description}, 
\begin{equation}\label{Psi_k}
\Psi_k=\left(\begin{array}{c}L'-1\\ k-1\end{array}\right)\left(\frac{G-B}{M+S-B}\right)^{k-1}\left(1-\frac{G-B}{M+S-B}\right)^{L'-k}.
\end{equation}
Applying (\ref{Psi_k}) to the definition of $\Psi(x)$,
%\footnotesize
\begin{eqnarray*}
\Psi(x) &=& \sum_{k=1}^{L'}\left(\begin{array}{c}L'-1\\ k-1\end{array}\right)\left(\frac{G-B}{M+S-B}\right)^{k-1}\left(1-\frac{G-B}{M+S-B}\right)^{L'-k}x^{k}\\
        &=& x\sum_{k=0}^{L'-1}\left(\begin{array}{c}L'-1\\ k\end{array}\right)\left(\frac{G-B}{M+S-B}x\right)^{k}\left(1-\frac{G-B}{M+S-B}\right)^{L'-1-k}\\
        &=& x\left[1-\frac{(G/B-1)(1-x)}{L'-1}\right]^{L'-1}\\
        &\approx& x\mathrm{e}^{-(G/B-1)(1-x)},
\end{eqnarray*}
where the approximation $\lim_{m\rightarrow\infty}(1+1/m)^m=\mathrm{e}$ is used, which is accurate even when $L'$ is not very large (e.g. $100$).

Let $\lambda(x)=\sum_{k=1}^{L'}\lambda_{k}x^{k-1}$ denote the left-side edge degree distribution, where $\lambda_k$ denotes the probability that a randomly chosen edge of the bipartite graph is connected with a left node of degree $k$. We have $\lambda(x)=\Psi'(x)/\Psi'(1)$ where $\Psi'(x)$ is the derivative of $\Psi(x)$ with respect to $x$. For sufficiently large $M$ and $L'$, we have 
\begin{equation}\label{lambda_asym}
\lambda(x) \approx \left(\frac{B}{G}+\left(1-\frac{B}{G}\right)x\right)\mathrm{e}^{-(G/B-1)(1-x)}.
\end{equation}

Suppose that each generation (i.e., right node) has received a number of linearly independent packets. The G-by-G decoding corresponds to the following process on the bipartite graph: initially, every node on the graph is labeled as \textit{unknown}, where we designate an edge as unknown if both of its end nodes are unknown, or \textit{known} if at least one of its end nodes is known. In each step, a right node is decoded by GE if it is \textit{full rank}, i.e., the number of its received linearly independent packets is larger than or equal to the number of its unknown adjacent edges. After decoding a right node, the decoder labels all adjacent edges of the decoded right node, its left neighbors, and edges adjacent to these neighbors as known. The decoding proceeds until no decodable right node can be found.

\begin{figure}[h]
\centerline{\includegraphics[height=1.5in]{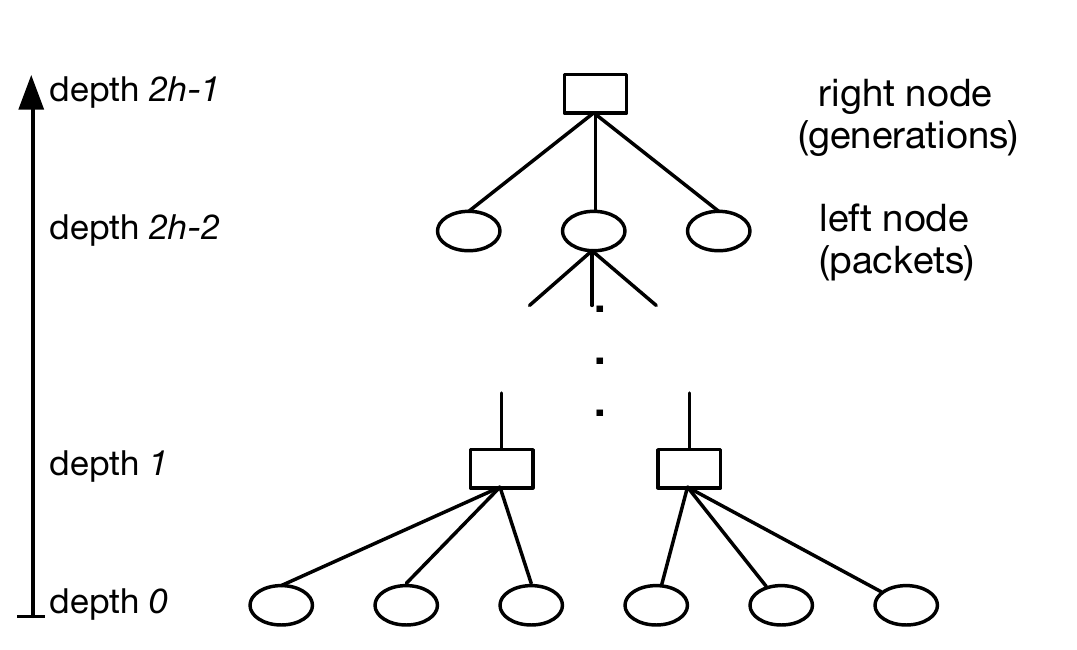}}
\caption{\label{tree}Expanding the graph as a tree.}
\end{figure}

The and-or tree evaluation technique of \cite{Luby1998} can be used to characterize the above decoding process by randomly choosing one edge of the bipartite graph and expanding the graph from the right node of the edge to obtain a tree. $h$ steps of G-by-G decoding correspond to the evaluation from depth-$0$ of a tree of depth $2h$ as shown in Fig. \ref{tree}. Nodes at depths $2h-1$ and $2h-2$ are referred to as on level $h$. 

Let $y_h$ and $z_h$ denote the probabilities that a right and left node on level $h$ are evaluated as unknown (i.e., not decoded), respectively. Then
\begin{equation}
y_h=\sum_{u=0}^{G-1}\eta_{u}\sum_{k=u}^{G-1}\left(\begin{array}{c}G-1\\ k\end{array}\right)(z_h)^k(1-z_h)^{G-1-k},
\end{equation}
where $\eta_{u}$ denotes the probability that $u$ linearly independent packets are received by a right node. Since $z_h=\lambda_1+\sum_{k\geq 2}\lambda_ky_{h-1}^{k-1}=\lambda(y_{h-1})$ where $z_0=1$, we have
\begin{equation}
y_h=\sum_{u=0}^{G-1}\eta_{u}\sum_{k=u}^{G-1}\left(\begin{array}{c}G-1\\ k\end{array}\right)(\lambda(y_{h-1}))^k(1-\lambda(y_{h-1}))^{G-1-k},
\end{equation}
which expresses the probability that a right node is unknown after one step of G-by-G decoding as a function of the corresponding probability before the step.

We denote the probability that a generation is not decodable as $x$, $x\in[\delta,1]$ where $\delta=\lim_{h\rightarrow\infty}y_h$ stands for the smallest probability that the decoder can reach after going through all generations. We therefore require 
\begin{equation*}
\sum_{u=0}^{G-1}\eta_{u}\sum_{k=u}^{G-1}\left(\begin{array}{c}G-1\\ k\end{array}\right)(\lambda(x))^k(1-\lambda(x))^{G-1-k}<x
\end{equation*}
for all $x\in[\delta,1]$ if we want at least a fraction $(1-\delta)$ of generations to be decoded, establishing (\ref{iterative_governer}). The inequality corresponds to the fact that the probability that a generation is not decodable is strictly decreasing.
\bibliographystyle{IEEEtran}
\bibliography{bibliography}
\end{document}